\documentclass[reprint,showpacs,aps,pra,amsmath,amssymb,floatfix,twocolumn, superscriptaddress]{revtex4-2}

\usepackage{graphicx}
\usepackage{siunitx}
\usepackage{physics}
\usepackage{MnSymbol}
\usepackage{mathtools}

\newcommand{\eqdef}{=\vcentcolon}
\usepackage{hyperref}
\hypersetup{
    colorlinks=true,
    citecolor=blue,
    linkcolor=blue,
    anchorcolor=black,
    urlcolor=blue,
}
\usepackage{csquotes}
\usepackage{bm}
\usepackage{xcolor}

\usepackage{natbib}

\begin{document}
\title{Magic running- and standing-wave optical traps for Rydberg atoms}
\author{Lukas Ahlheit}
\affiliation{Institut f\"{u}r Angewandte Physik, University of Bonn, Wegelerstr. 8, 53115 Bonn, Germany}
\author{Chris Nill}
\affiliation{Institut f\"{u}r Angewandte Physik, University of Bonn, Wegelerstr. 8, 53115 Bonn, Germany}
\affiliation{Institut f\"{u}r Theoretische Physik and Center for Integrated Quantum Science and Technology, Universität T\"{u}bingen, Auf der Morgenstelle 14, 72076 T\"{u}bingen, Germany}
\author{Daniil Svirskiy}
\affiliation{Institut f\"{u}r Angewandte Physik, University of Bonn, Wegelerstr. 8, 53115 Bonn, Germany}
\author{Jan de Haan}
\affiliation{Institut f\"{u}r Angewandte Physik, University of Bonn, Wegelerstr. 8, 53115 Bonn, Germany}
\author{Simon Schroers}
\affiliation{Institut f\"{u}r Angewandte Physik, University of Bonn, Wegelerstr. 8, 53115 Bonn, Germany}
\author{Wolfgang Alt}
\affiliation{Institut f\"{u}r Angewandte Physik, University of Bonn, Wegelerstr. 8, 53115 Bonn, Germany}
\author{Nina Stiesdal}
\affiliation{Institut f\"{u}r Angewandte Physik, University of Bonn, Wegelerstr. 8, 53115 Bonn, Germany}
\author{Igor Lesanovsky}
\affiliation{Institut f\"{u}r Theoretische Physik and Center for Integrated Quantum Science and Technology, Universität T\"{u}bingen, Auf der Morgenstelle 14, 72076 T\"{u}bingen, Germany}
\affiliation{School of Physics and Astronomy and Centre for the Mathematics and Theoretical Physics of Quantum Non-Equilibrium Systems, The University of Nottingham, Nottingham, NG7 2RD, United Kingdom}
\author{Sebastian Hofferberth}
\affiliation{Institut f\"{u}r Angewandte Physik, University of Bonn, Wegelerstr. 8, 53115 Bonn, Germany}

\date{\today}

\begin{abstract}
Magic trapping of ground and Rydberg states, which equalizes the AC Stark shifts of these two levels, enables increased ground-to-Rydberg state coherence times.
We measure via photon storage and retrieval how the ground-to-Rydberg state coherence depends on trap wavelength for two different traps and find different optimal wavelengths for a one-dimensional optical lattice trap and a running wave optical dipole trap. Comparison to theory reveals that this is caused by the Rydberg electron sampling different potential landscapes.
The observed difference increases for higher principal quantum numbers, where the extent of the Rydberg electron wave function becomes larger than the optical lattice period.
Our analysis shows that optimal magic trapping conditions depend on the trap geometry, in particular for optical lattices and tweezers.
\end{abstract}

\maketitle

\section{Introduction}
Converting photons into propagating or stationary coherent excitations in atomic ensembles is a key aspect to quantum memory schemes
~\cite{harris1991,lukin2001,hau2001,pan2009,Tittel2009,pu2024}.
Photon storage times in atomic media are limited by various decoherence mechanisms, including thermal atomic motion, which can be reduced by using ultracold atoms, and inhomogeneous differential light shifts between atoms sharing the coherence~\cite{you2003, vuletic2007, pan2009, durr2020}.
The latter can be reduced by turning off optical traps or by using magic wavelength traps where both involved atomic states experience the same light shift~\cite{kennedy2010}.
This approach is common for quantum memories based on hyperfine ground state transitions and enables photon storage on minute time-scales~\cite{kuzmich2013}.

Slow light and photon storage schemes involving one or more Rydberg states enable mapping of strong atomic interactions onto propagating or stored photons~\cite{lukin2000,zoller2001,vuletic2012,kuzmich2012,vuletic2013,adams2013,hofferberth2016b,adams2017,firstenberg2023}.
Compared to transitions between hyperfine ground states, one challenge of using Rydberg states is that the two-photon excitation scheme usually results in large $k$-vector mismatch between probe and control light, leading to a faster loss of coherence with faster atomic motion~\cite{pan2009}.
This motional dephasing can be prevented by trapping the atoms in a standing-wave optical lattice. The extend of the Rydberg electron wave function, comparable to the lattice period in typical optical traps, requires careful calculation of the magic condition.
The quivering motion of the almost-free Rydberg electron in the electric trapping field leads to a repulsive ponderomotive potential~\cite{raithel2000,r.-s.-harvie1957,batelaan2001}, which gives rise to a repulsive optical potential for the entire Rydberg atom~\cite{walker2005}.
One solution is an effectively blue-detuned trap for both ground state and Rydberg state~\cite{lahaye2020,Sayrin2023,derevianko2013a,meinert2024}.
Another approach is to use a red-detuned trap for the ground state which at the same time near-resonantly couples the Rydberg state to a lower lying atomic state~\cite{walker2005, safronova2015, kuzmich2018} to overcome the repulsive ponderomotive potential.
Such near-resonant traps with minimized differential AC Stark shift have been employed in an optical lattice to increase ground-to-Rydberg state coherence to approximately $\SI{20}{\micro s}$, approaching the Rydberg state lifetime~\cite{kuzmich2018}.
In such an optical lattice, the Rydberg electron samples the potential landscape surrounding the atomic core. Therefore, the trapping geometry will influence the exact magic wavelength when the size of the Rydberg electron wave function gets comparable to the trap dimensions~\cite{raithel2000}.

In this paper, we explore this geometry dependence in detail by comparing magic trapping wavelengths for atoms in a one-dimensional optical lattice and a single-beam running wave trap.
For experimentally determining the magic trapping conditions, we measure the coherence time of a collective ground-state-Rydberg-state excitation as a function of trapping wavelengths.
We review ground and Rydberg state potentials and derive magic trapping conditions as function of trap laser wavelength and Rydberg state principal quantum number $n$.
Our experiments show that both the longitudinal standing wave as well as the radial trap shape influence the magic condition.
We find that the experimentally best magic condition is obtained by minimizing the variance of the differential light shift across the atomic ensemble.
We verify the modification of optimal wavelengths for lattice trapped atoms for the increasing Rydberg electron wave function with $n$.

\section{Experimental Setup}
\label{sec:experiment_setup}
\begin{figure*}
    \includegraphics[width=\textwidth]{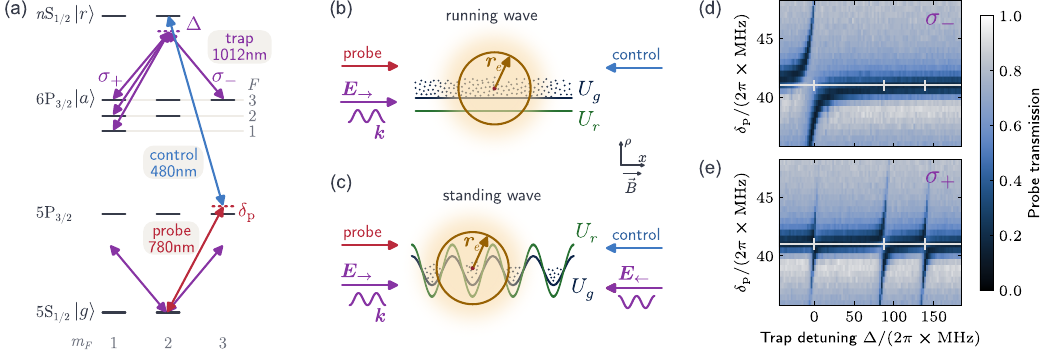}
    \caption{Setup for the magic wavelength trap.
    (a)~Energy level diagram of relevant $^{87}\mathrm{Rb}$ hyperfine states. The red and blue arrows indicate the two-photon excitation (probe, control) from the ground state $\ket{g}$ to the Rydberg state $\ket{r}$.
    The purple arrows show possible couplings between the Rydberg state $\ket{r}$ and different $6P_{3/2}$ hyperfine states by the circularly polarized trapping laser light. The detuning $\Delta$ is the one of the trapping laser from the transition between the $6P_{3/2}$ hyperfine state $\ket{a}$ and the Rydberg state $\ket{r}$.
    (b),~(c)~The used co-linear optical beams in the two trapping configurations.
    We compare a running wave trap (b), made of an incident beam with electric field $\bm{E}_\rightarrow$, with a standing wave trap (c), formed by adding a counter-propagating beam with electric field $\bm{E}_\leftarrow$. The two-photon excitation beams counter-propagate along the trap axis. The magnetic field points along the probe beam to define the quantization axis. The blue and green lines illustrate the trapping potential experienced by ground state and Rydberg atoms, respectively. 
    The case sketched here is for a non-magic trap detuning where the two different states experience different trapping potentials.
    $r_e$ indicates the extent of the Rydberg electron wave function for $76S$.
    (d),~(e)~Experimentally measured probe laser transmission spectroscopy of the two-photon transition to $108S$ for varying the probe detuning $\delta_\mathrm{p}$ and the trap laser detuning $\Delta$. The control field is detuned by $2\pi \times \SI{41}{MHz}$ (white horizontal line) relative to the bare $\ket{5P_{3/2},F=3,m_F=3}-\ket{r}$-transition.
    When the trapping beam drives $\sigma^-$ (d) or $\sigma^+$ (e) transitions, we observe one or three hyperfine transitions, respectively.
    }
    \label{fig1:cartoon_and_levels}
\end{figure*}

Fig.~\ref{fig1:cartoon_and_levels}(a) shows the relevant level scheme of $^{87}$Rb and the involved lasers for magic trapping and Rydberg excitation. 
To create attractive potentials for both the ground and the Rydberg state, we employ a $\lambda_T\approx\SI{1012}{\nano\meter}$ trap laser. 
This laser creates a far-off resonant dipole trap for atoms in the ground state $|g\rangle = \ket{5S_{1/2},F=2,m_F=2}$.
To also trap atoms in the Rydberg state $\ket{r}=\ket{nS_{1/2},m_J=1/2}$, it near-resonantly couples the Rydberg state to the lower-lying $\ket{6P_{3/2}}$ state.
As indicated in the level scheme, the trap can couple $\ket{r}$ to different $6P_{3/2}$ hyperfine states depending on its polarization. We use $\sigma^-$ polarized trapping light to couple only to a single, well-defined hyperfine state $\ket{a}=\ket{6P_{3/2},F=3,m_F=3}$.
We define the trap detuning as $\Delta=\omega-\omega_{ra}$, where $\omega$ is the laser frequency and $\omega_{ra}$ is the transition frequency between $\ket{r}$ and $\ket{a}$.
To create the collective ground-state-Rydberg-state excitation, we use the two-photon excitation scheme shown in Fig.~\ref{fig1:cartoon_and_levels}(a).
Few-photon probe pulses of $\SI{780}{nm}$ together with a strong $\SI{480}{nm}$ control field couple $|g\rangle$ via the intermediate state $\ket{5P_{3/2},F=3,m_F=3}$ to $\ket{r}$. Both beams are focused onto the trapped atoms.
The probe beam has a $1/e^2$ intensity waist radius of $\SI{5.1}{\micro\meter}$ and the counter-propagating control beam has a waist radius of $\SI{45.1}{\micro\meter}$.

Fig.~\ref{fig1:cartoon_and_levels}(b),(c) show the two different trap geometries where we investigate magic trapping of Rydberg atoms.
The probe and control beams are co-linear with the trapping beams.
Fig.~\ref{fig1:cartoon_and_levels}(b) shows the running wave trap that is created by a single trapping beam with electric field~$\bm{E}_\rightarrow$ focused at the atom cloud center with a $1/e^2$ radius of $\SI{24.8}{\micro\meter}$.
In Fig.~\ref{fig1:cartoon_and_levels}(c) this single beam is retro-reflected into electric field $\bm{E}_\leftarrow$ to form a standing-wave lattice. 
Throughout this paper, we set the incoming power in the standing wave to be one fourth of the power used for the running wave in order to have the same maximum intensity in both geometries.
The resulting ground state central trap depth is $\SI{12}{\micro K \cdot \textit{k}_{B}}$.
We apply a magnetic field parallel to the probe beam, along which we also define the quantization axis of the system.
We load ultracold $^{87}\mathrm{Rb}$ atoms at a temperature of $T \sim\SI{2}{\micro K}$ into either of the two geometries.
We adjust the number of atoms in the final trap between $500$ and $5000$ atoms depending on the target Rydberg state~\footnote{We adjust the final atom number by varying the initial magnetic-optical trap loading parameters.
For $n>70$ we use lower atomic densities to reduce the formation of Rydberg molecules as discussed later in the main text.
As we vary $n$ we thereby also reduce density dependent dephasing~\cite{durr2014,pfau2014a}.}.
The axial dimension of the atomic distribution is determined by the initial atom preparation and has a $1/e^2$ half-width of $\SI{40.8}{\micro\meter}$.
In the radial direction, the atom distribution is defined by the investigated trap geometry and has a $1/e^2$ half-width of $\SI{13.4}{\micro\meter}$.

In Fig.~\ref{fig1:cartoon_and_levels}(d), (e) we show the probe transmission spectroscopy to $\ket{r}=108S_{1/2}$ in the presence of the trap laser as a function of probe detuning $\delta_p$ and trap detuning $\Delta$.
In Fig.~\ref{fig1:cartoon_and_levels}(d) the trapping field drives $\sigma^-$ transitions. At $\Delta=\SI{0}{MHz}$, the coupling between $\ket{r}$ and $\ket{6P_{3/2},F=3,m_F=3}$ results in an avoided crossing.
The white horizontal line marks the ground-to-Rydberg transition measured in free space.
In Fig.~\ref{fig1:cartoon_and_levels}(e) the trap drives $\sigma^+$ transitions, and avoided crossings appear for $\ket{6P_{3/2},F=1,2,3,m_F=1}$.
The observed level spacing is in agreement with known $6P_{3/2}$ hyperfine splitting~\cite{fortagh2020}, which is indicated by white ticks.
Our combination of polarizations allows us to realize an isolated $4$-level system with three beams propagating parallel to the quantization axis.
We avoid any rotation of optically pumped spin-orientation in the experiment sequence, open transitions~\cite{kuzmich2019} or perpendicular beams~\cite{kuzmich2013a}.

\section{Theory: trapping potential}
\label{sec:theory}
In the following, we review the theory of Rydberg atom optical trapping, which in particular has to take into account that the Rydberg wave function is large compared to the optical wavelength~\cite{raithel2000}. Our derivation follows the one given by Lampen~\textit{et al.}~\cite{kuzmich2018}.
For each geometry we derive the trapping potentials $U_g$ and $U_r$ experienced by an atom in the ground and the Rydberg state respectively.
The trapping laser, at position $\bm{r}$ and time $t$, is generally characterized by the vector potential $\bm{A}(\bm{r},t)$. In the Coulomb gauge, the resulting electric field is given by $\bm{E}(\bm{r},t)=-\partial \bm{A}(\bm{r},t)/\partial t$.
We consider the atom core to be located at position~$\bm{r}$ and the single valence electron at~$\bm{r}+\bm{r}_\mathrm{e}$ in the core potential~$V_\mathrm{c}(\bm{r}_\mathrm e)$.
For both ground and Rydberg state, the Hamiltonian is given by the minimal-coupling Hamiltonian of an electron in an electromagnetic field
\begin{align}
    H
    &=\frac{1}{2m_\mathrm{e}} \left(\bm{p}_\mathrm{e}-e\bm{A}(\bm{r}+\bm{r}_\mathrm{e},t)\right)^2+eV_\mathrm{c}(\bm{r}_\mathrm{e})\label{eq:general-Hamiltonian}\\
    &=\underbrace{\frac{\bm{p}_\mathrm{e}^2}{2m_\mathrm{e}}+eV_\mathrm{c}(\bm{r}_\mathrm{e})}_{H_0}
    \underbrace{- \frac{e\bm{A}(\bm{r}+\bm{r}_\mathrm{e},t) \cdot \bm{p}_\mathrm{e}}{m_\mathrm{e}}}_{H_1}+\underbrace{\frac{e^2\bm{A}^2(\bm{r}+\bm{r}_\mathrm{e},t)}{2m_\mathrm{e}}}_{H_2}.\nonumber
\end{align}
Here, $e$ and $m_\mathrm{e}$ are the electron charge and mass, and $\bm{p}_\mathrm{e}$ the momentum operator of the electron.
The Hamiltonian (second row of Eq.~(\ref{eq:general-Hamiltonian})) is now in the form of a hydrogen-like Hamiltonian~$H_0$ perturbed by~$H_{1,2}$.
For an eigenstate $\ket q$ of $H_0$, the energy is given as $\hbar\omega_q\ket q=H_0\ket q$.
We calculate the energy shift of both ground and Rydberg states by treating~$H_{1,2}$ in perturbation theory.
Therefore, the time dependence of $H_{1,2}$ has to be decomposed in Floquet-modes using the Floquet-quasi-energy approach~\cite{krainov1985,juzeliunas2017}.
Hence, a concrete form of the vector potential is needed.

For two counter propagating Gaussian beams, as we use to form our trap configuration, the vector potential is given by
\begin{align}
    \bm{A}(\bm{r}+\bm{r}_\mathrm{e},t)=\frac{\bm{\epsilon}}{\omega} (
    &E_\rightarrow(\bm{r}) \sin{\left[(\bm{r}+\bm{r}_\mathrm{e})\cdot\bm{k} - \omega t \right]}\nonumber\\
    +&E_\leftarrow(\bm{r}) \sin{\left[(\bm{r}+\bm{r}_\mathrm{e})\cdot\bm{k} + \omega t \right]}),
    \label{eq:vector-potential}
\end{align}
with the polarization $\bm{\epsilon}$ and the wave vector $\bm{k}=k\bm{\hat{e}}_x$ of the trap laser propagating along the $x$-axis.
For a running wave configuration, the electric field amplitude in backwards direction is $E_\leftarrow(\bm{r})=0$.
Note that we assume that the Gaussian-shaped field amplitude is homogeneous over the diameter of the Rydberg (ground state) atom. Therefore, we set $E_{\rightarrow(\leftarrow)}(\bm{r}+\bm{r}_\mathrm{e})\approx E_{\rightarrow(\leftarrow)}(\bm{r})$.

First, we calculate the energy shift caused by $H_1$ acting on an eigenstate $\ket q$ of the unperturbed Hamiltonian.
A detailed derivation is outlined in Appendix~(\ref{appendix:H1-correction}).
We first decompose $H_1$ in Floquet-modes as
\begin{align}
    H_1=\frac{e\bm{A}(\bm{r}+\bm{r}_\mathrm e)\cdot\bm p_\mathrm e}{m_\mathrm e}e^{i\omega t}+\frac{e\bm{A}^*(\bm{r}+\bm{r}_\mathrm e)\cdot\bm p_\mathrm e}{m_\mathrm e}e^{-i\omega t},
\end{align}
see Appendix~(\ref{appendix:vector-potential}) for the expanded form of the vector potential in Eq.~(\ref{eq:A-e-iomegat}).
For the energy correction in second order, we then obtain
\begin{align}
    U_q^{(1)}(\bm{r})
    &\approx\abs{\bm{A}(\bm{r})}^2\sum_{j\neq q}
    \frac{\omega_{qj}^2}{\hbar}\left(
    \frac{\abs{\langle q|\bm\epsilon \cdot \bm{r}_\mathrm e|j\rangle}^2}{\omega_{qj}-\omega}
    +\frac{\abs{\langle q|\bm \epsilon^* \cdot \bm{r}_\mathrm e|j\rangle}^2}{\omega_{qj}+\omega}\right),
    \label{eq:U_1}
\end{align}
where $\omega_{qj}=\omega_q-\omega_j$.

Thus, we obtain the dynamic polarizability $\alpha_g^{(\mathrm b)}$ arising from the bound-state transitions in
\begin{equation}
    U_g^{(1)}(\bm{r})\approx-\alpha_g^{(\mathrm b)} \omega^2{\abs{\bm{A}(\bm{r})}^2}.
    \label{eq:U-g-1}
\end{equation}

For the energy shift of the Rydberg state $U_r^{(1)}$, the resonance approximation can be made. This is possible since the electromagnetic field with frequency $\omega$ is only in near resonance to the transition from $\ket a$ to $\ket r$.
Therefore, other transitions are negligible, and the summation in Eq.~(\ref{eq:U_1}) gets drastically simplified. With $\Delta=\omega-\omega_{ra}$ and our polarization, we obtain
\begin{align}
    U_r^{(1)}
    &\approx -\frac{D_{ar}^2}{8\hbar\Delta}
    \omega^2{\abs{\bm{A}(\bm{r})}}^2,
    \label{eq:U_r1}
\end{align}
with the reduced dipole matrix element $D_{ar}=\mel{a}{\abs{e\bm{r}_\mathrm{e}}}{r}$.

The energy shift arising from $H_2$ is non-vanishing in first order perturbation theory. Therefore, we decompose $H_2$ in Floquet modes, see Eq.~(\ref{eq:A2-e-2iomegat}).
The first-order energy shift of state $\ket q$ only depends on Floquet mode~$0$, thus
\begin{equation}
     U_q^{(2)}\approx \frac{e^2}{2m_\mathrm e}\langle q|2 \bm{A}(\bm{r}+\bm{r}_\mathrm e)\bm{A}^*(\bm{r}+\bm{r}_\mathrm e)|q\rangle.
\end{equation}
 
We explicitly evaluate $U_q^{(2)}$ by using the expanded form of the vector potential (see Eq.~(\ref{eq:vector-potential-avg-squared})) and obtain
\begin{align}
     U_q^{(2)}
     \approx-\frac{\alpha_f}{4}\Big[&4E_\rightarrow (\bm{r})E_\leftarrow (\bm{r})\cos^2(kx)\theta_q\nonumber\\
     &-2 E_\rightarrow (\bm{r})E_\leftarrow (\bm{r})\sin(2kx)\ev{
    \sin(2kx_\mathrm e) }{q}\nonumber\\
    &+2E_\rightarrow (\bm{r})E_\leftarrow(\bm{r})
    \left[1-\theta_q\right]\nonumber\\
    &+\Big(E_\rightarrow (\bm{r})- E_\leftarrow (\bm{r}) \Big)^{2}\Big],
\end{align}
where $\alpha_f=-e^2/m_\mathrm{e}\omega^2$ is the free-electron polarizability.
Since $\sin(2kx_\mathrm e)$ is an odd function, $\ev{\sin(2kx_\mathrm e)}{q}=0$ due to parity arguments.
We introduce the landscape factor $\theta_q=\ev{\cos(2kx_\mathrm e)}{q}$, to describe the electron wave function sampling the variation of the optical lattice potential~\cite{derevianko2013,kuzmich2018}.
This term becomes significant when the extension of the radial wave function becomes comparable to the optical trap period.
In this case, the optical dipole approximation breaks down.
However, for the ground state the dipole approximation with $kx_\mathrm e\ll1$ can be applied and thus $\theta_g=\ev{\cos(2kx_\mathrm e)}{g}\approx 1$.

In contrast, for the Rydberg state, the landscape factor $\theta_r$ has to be calculated explicitly.
In our case, $\ket{r} = \ket{nS_{1/2}, m_J = 1/2}$ is an S-state.
To calculate the landscape factor as function of $n$, we use spherical Bessel functions as
\begin{equation}
    \theta_n = \ev{\cos(2kx_\mathrm{e})}{nS} = \int_{0}^{\infty} dr_e P_{n,0}(r_e)j_0(2kr_e),
    \label{eq:theta_n approx}
\end{equation}
where $P_{n,0}(r_e)$ is the radial probability density of an atom in state $\ket{nS}$, and $j_0(2kr_e)$ is the spherical Bessel function of the first kind for zero angular momentum~\cite{kuzmich2018}.

Summarizing the derived terms, we obtain the energy shift for ground state atoms as
\begin{align}
    U_g(\bm{r})&\approx U_g^{(1)}(\bm{r})+U_g^{(2)}(\bm{r})\\
    &=-( \underbrace{\alpha_g^{(\mathrm b)}+\alpha_f}_{\alpha_g}) \omega^2{\abs{\bm{A}(\bm{r})}^2}.
\end{align}
Here, $\alpha_g$ is the well-known dynamic polarizability of the ground state atom~\cite{derevianko2013,ovchinnikov2000, drake2023a}.
For the Rydberg state, we obtain
\begin{align}
U_r(\bm{r})&\approx U_r^{(1)}(\bm{r})+U_r^{(2)}(\bm{r})\\
    &= -\frac{D_{ar}^2}{8\hbar\Delta}
    \omega^2{\abs{\bm{A}(\bm{r})}^2}+U_r^{(2)}(\bm{r}).
\end{align}

We split the potentials into the axially periodic potential $U_g^\sim(\bm{r})$, denoted with $\sim$, and the axially non-periodic potential $U_g^-(\bm{r})$, denoted as --, thus
\begin{align}
    U_g(\bm{r})&=U_g^{\sim}(\bm{r})+U_g^{-}(\bm{r}),\label{eq:U_g}\\
    U_g^{\sim}(\bm{r}) &= -\alpha_g E_\rightarrow(\bm{r}) E_\leftarrow(\bm{r}) \cos ^2(kx),\label{eq:U_g_periodic}\\
    U_g^{-}(\bm{r}) &= -\frac{\alpha_g}{4}\Big(E_\rightarrow(\bm{r})-E_\leftarrow(\bm{r})\Big)^2.\label{eq:U_g_nonperiodic}
\end{align}
For numerical calculations, we use matrix elements provided by the ARC package~\cite{jones2021}.
For the Rydberg state, we obtain
\begin{align}
    U_r(\bm{r})=& U_r^\sim(\bm{r})+U_r^-(\bm{r}),\label{eq:U_r}\\
    U_r^{\sim}(\bm{r})=& - E_\rightarrow(\bm{r}) E_\leftarrow(\bm{r}) \cos^2\left(k x\right)
        \left(\frac{D_{ar}^2}{4 \hbar \Delta}+\alpha_f \theta_n\right),\label{eq:U_r_periodic}\\
    U_r^-(\bm{r}) =& -\frac{D_{ar}^2}{16\hbar \Delta}\left[E_\rightarrow(\bm{r})-E_\leftarrow(\bm{r})\right]^2 \nonumber\label{eq:U_r_nonperiodic}\\
        & -\frac{\alpha_f}{4}\big[2 E_\rightarrow(\bm{r}) E_\leftarrow(\bm{r})\left(1-\theta_n\right)\nonumber \\
        &+ \left[E_\rightarrow(\bm{r})-E_\leftarrow(\bm{r})\right]^2\big].
\end{align}
The light shifts of the ground and Rydberg state correspond to our measured avoided crossing in Fig.~\ref{fig1:cartoon_and_levels}(d), (e).
In the rest of the paper, we decompose positions $\bm{r}$ into the radial position $\rho$ and the axial position $x$.

\subsection*{Determining the magic detuning}
\begin{figure}
    \centering
    \includegraphics[width=\linewidth]{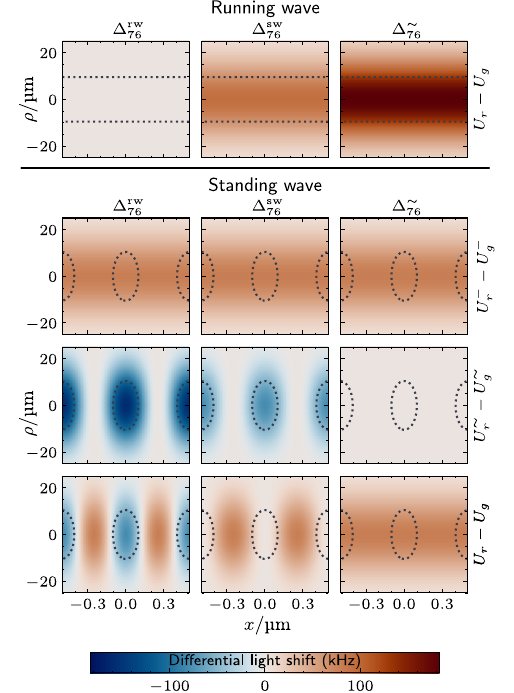}
    \caption{Calculated differential light shift between ground state and Rydberg state $76S$ for running- and standing-wave configuration with varying trap detunings $\Delta$ (columns).
    The light shifts from Eqs.~(\ref{eq:U_g})-(\ref{eq:U_r_nonperiodic}) are evaluated for radial position $\rho$ and axial position $x$ relative to the trap center.
    The dotted lines indicate the atomic distribution, see main text.
    For a running wave (first row), the differential light shift vanishes everywhere at the detuning $\Delta^\mathrm{rw}_{76}/2\pi=336\,$MHz (left column).
    The potential of the standing wave trap can be decomposed into a non-periodic (--) (second row) and a periodic ($\sim$) (third row) component.
    For detuning $\Delta^\sim_{76}/2\pi=548\,$MHz (right column), the periodic light shift vanishes, while a non-periodic component remains.
    For our experimental parameters, the full standing-wave trap (fourth row) shows an optimal detuning around $\Delta^\mathrm{sw}_{76}/2\pi=413\,$MHz (center column), where the standard deviation of the light shift over the whole atom cloud is minimal (see Fig. \ref{fig3:std_vs_det}).
    }
    \label{fig2:potential_2D_maps}
\end{figure}

We can now calculate the differential light shift $U_r(\rho,x,\Delta) - U_g(\rho,x)$ using Eqs.~(\ref{eq:U_g})-(\ref{eq:U_r_nonperiodic}). In Fig.~\ref{fig2:potential_2D_maps} we show the differential light shift for the $76S$ Rydberg state for our two different trap geometries.
The first row shows the potential for the running wave ($E_\leftarrow=0$), and the second, third and fourth row illustrate the standing-wave potential ($E_\leftarrow=0.87 \cdot E_\rightarrow$, limited by the power loss over the retro-reflection optical path).
To illustrate the extent of the atomic cloud in the trap, the dashed contour lines show the position where the ground state trapping potential~$U_g(\rho,x)$ equals two times the atom thermal energy~$k_B T$.
The first row shows the differential light shift for a running wave trap.
The trap detuning $\Delta^\text{rw}_n$ that satisfies $U_r(\rho,x,\Delta^\text{rw}_n)-U_g(\rho,x)=0$ is found analytically:
\begin{equation}
    \frac{D_{ar}^2}{4\hbar\Delta^\text{rw}_n} = \alpha_g - \alpha_f.
    \label{eq:magic condition running wave}
\end{equation}
The change in differential light shift as function of trap laser detuning $\Delta$ comes solely from the Rydberg state potential because of the near-resonant coupling to the $\ket{a} \rightarrow \ket{r}$ transition.

The second and third row in Fig.~\ref{fig2:potential_2D_maps} separately show the two contributions to the differential shift in the standing wave, as introduced in Eqs.~(\ref{eq:U_g})-(\ref{eq:U_r_nonperiodic}).
The second row shows the differential light shift from the non-periodic components $U_{(g,r)}^-$. For our imbalanced standing wave, this differential light shift component only weakly depends on the trap detuning $\Delta$.
In the third row, the differential light shift from the periodic components $U_{(g,r)}^\sim$ is shown.
We find the detuning $\Delta^\sim_n$ that satisfies $U_r^{\sim}(\rho,x,\Delta^\sim_n)-U_g^{\sim}(\rho,x)=0$ analytically by
\begin{equation}
    \frac{D_{ar}^2}{4\hbar\Delta^\sim_{n}} = \alpha_g - \alpha_f \theta_n.
    \label{eq:magic condition periodic}
\end{equation}
The third column shows the potential for $\Delta = \Delta^\sim_{76}$, where the differential light shift from the periodic components is canceled.
The differential potential in a standing wave is given by the sum of the periodic and the non-periodic components, and is shown in the fourth row of Fig.~\ref{fig2:potential_2D_maps}.
The periodic and non-periodic components of the full standing wave do not have the same radial and axial dependence, and therefore it is not possible to find a single, position-independent magic detuning. Instead, we consider the detuning that minimizes the inhomogeneity of the differential light shift across the atomic cloud.

We quantify this by calculating the standard deviation of the light shift over the active volume of the atomic cloud.
This volume is determined by both the atomic cloud shape and the excitation beams.
We consider the cloud volume inside $U_g(\rho,x)\leq U_g(0,0)+2\cdot k_B T$ and two times the $1/e^2$ radius of our probe beam.
The middle column of Fig.~\ref{fig2:potential_2D_maps} shows the differential light shift for the detuning $\Delta^\mathrm{sw}_{76}$ where the standard deviation is minimal.

\begin{figure}
    \centering
    \includegraphics[width=\linewidth]{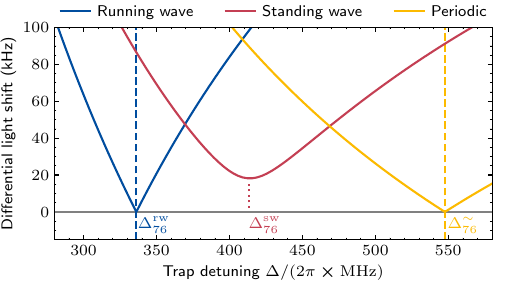}
    \caption{Standard deviation of the differential light shift between ground state and Rydberg state $76S$ evaluated inside the dotted volumes highlighted in Fig.~\ref{fig2:potential_2D_maps} for our experimental parameters.
    The vertical dashed lines show $\Delta^\mathrm{rw}_{76}$ and $\Delta^\sim_{76}$ from Eqs.~(\ref{eq:magic condition running wave}) and (\ref{eq:magic condition periodic}). 
    The standard deviation of the differential light shift in a full standing wave shows an optimal trap detuning $\Delta^\mathrm{sw}_{76}$ between $\Delta^\mathrm{rw}_{76}$ and $\Delta^\sim_{76}$.
    }
    \label{fig3:std_vs_det}
\end{figure}

Fig.~\ref{fig3:std_vs_det} shows the standard deviation of the differential light shift as function of trap detuning for the running wave, the standing wave, as well as for the purely periodic potential component.
The minimum of the standard deviations in the running wave and the purely periodic potentials coincide with the calculated optimal detunings described by Eqs.~(\ref{eq:magic condition running wave}) and (\ref{eq:magic condition periodic}).
The optimum detuning for the full standing-wave trap lies between the magic detunings for the running wave and the periodic potential.
Neither of the limiting cases describe our implemented standing-wave trap well.

\section{Experimental results}
\label{sec:experimental_results}
We measure the trap-induced decoherence for single photons stored as collective excitations in the medium~\cite{lukin2000,lukin2001,hau2001,lukin2002,hofferberth2017}.
We read out the stored probe photons after a variable storage time $t_s$.
The retrieved photons are detected by fiber coupled single photon counting modules.
We normalize the time-dependent retrieval signal $\eta(t_s)$ to the value for the shortest storage time of $\SI{0.5}{\micro s}$.
The control laser is on-resonance with the $\ket{5P_{3/2},F=3,m_F=3}-\ket{r}$-transition during both the photon storage and retrieval.

\begin{figure}[t]
    \centering
    \includegraphics[width=\linewidth]{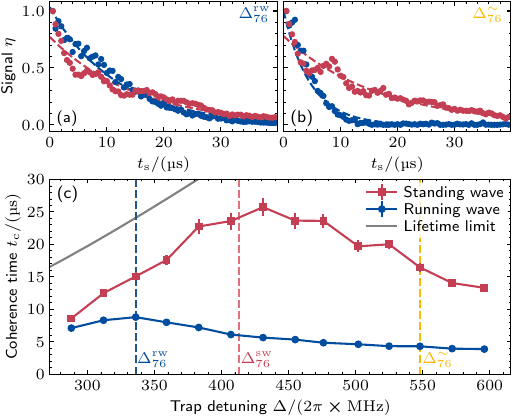}
    \caption{Photon storage measurements with Rydberg state $76S$ for different trap laser detunings. (a), (b) Normalized retrieved signal $\eta$ as a function of storage time $t_\mathrm{s}$ for atoms trapped in a standing wave (red) and running wave (blue). Panel (a) is measured for $\Delta_{76}^\mathrm{rw}$ and (b) at $\Delta_{76}^\sim$. Exponential fits of Eq.~(\ref{eq:exp-fit}) to the data, shown here as dashed lines, are used to extract a coherence time $t_c$. (c) The coherence time $t_c$ as function of trap laser detuning. The dashed vertical lines mark the analytic magic conditions for running wave Eq.~(\ref{eq:magic condition running wave}), and periodic potential Eq.~(\ref{eq:magic condition periodic}), as well as the optimal standing wave detuning $\Delta_{76}^\mathrm{sw}$.
    }
    \label{fig4:photon_storage_measurements}
\end{figure}

In Fig.~\ref{fig4:photon_storage_measurements}(a) and (b) we show $\eta(t_s)$ for $76S$ for the running wave and the standing-wave trap introduced in Fig.~\ref{fig1:cartoon_and_levels}.
The standing-wave measurements in Fig.~\ref{fig4:photon_storage_measurements}(a) and (b) show two additional oscillations on top of the exponential decay.
The oscillation with a period of $\SI{8}{\micro s}$ results from the excitation of Rydberg molecule states in the storage process.
These Rydberg dimers, made of a Rydberg atom and a ground state atom, form due to scattering between the Rydberg electron and the ground state atom~\cite{sadeghpour2000,pfau2009}.
Photons are stored in a superposition of collective Rydberg atom and dimer excitations, causing the beat signal on top of the retrieval curve.
The amplitude of the beat signal scales with the atomic density.
The Rydberg dimer excitation energy is shifted compared to the Rydberg atom transition by the dimer binding energy. For $n=76$ the binding energy is $\SI{114}{kHz}$~\cite{hofferberth2017}.
For atoms in the lattice, there is also a less prominent second oscillation with a $\SI{24}{\micro \second}$ period, which stems from the motion of the atoms in the lattice wells~\cite{kuzmich2018}. The observed frequency is close to the axial trap frequency, but deviates slightly due to the anharmonicity of the trapping potential.
With both of these oscillations present,
we use a simple exponential fit with the function
\begin{equation}
    f(t_s)=A \cdot e^{-t_s/t_c}
    \label{eq:exp-fit}
\end{equation}
to average through the oscillations and extract a model-independent coherence time $t_c$.
We expect to measure the longest coherence time at the optimal detuning calculated in Fig.~\ref{fig3:std_vs_det}.

In Fig.~\ref{fig4:photon_storage_measurements}(c) we show the fitted coherence time as a function of trap laser detuning $\Delta$.
The coherence time for the running wave peaks at the magic detuning $\Delta^\mathrm{rw}_{76}$, see Fig.~\ref{fig4:photon_storage_measurements}(a) and (b).
For the standing wave, we find a clear discrepancy from the pure periodic potential of Eq.~(\ref{eq:magic condition periodic}). Instead, the experimental result agrees well with the magic detuning $\Delta^\mathrm{sw}_{76}$ obtained by the standard deviation calculation shown in Fig.~\ref{fig3:std_vs_det}.
The optimal detuning will only approach the simple lattice calculation in the limiting case where atoms do not experience the radial variation of the trap potential.
The waist of the probe laser beam should be at least $20$ times smaller than the trap waist to reach this limit.

We calculate the coherence time limit due to admixture of the fast-decaying $6P_{3/2}$ state, which scales with~$\Delta^2$~\cite{kuzmich2018}, and show it as a gray line in Fig.~\ref{fig4:photon_storage_measurements}.
We find the same optimal detunings when we subtract this contribution to the coherence time from our measured data.
The decoherence caused by the differential light shift is the only significant detuning dependent mechanism for our experimental parameters.
Other effects like atomic motion and Rydberg-Rydberg-interaction generally limit the coherence time, but are independent of the trap detuning~\cite{durr2020}.

\begin{figure}[t]
    \centering
    \includegraphics[width=\linewidth]
    {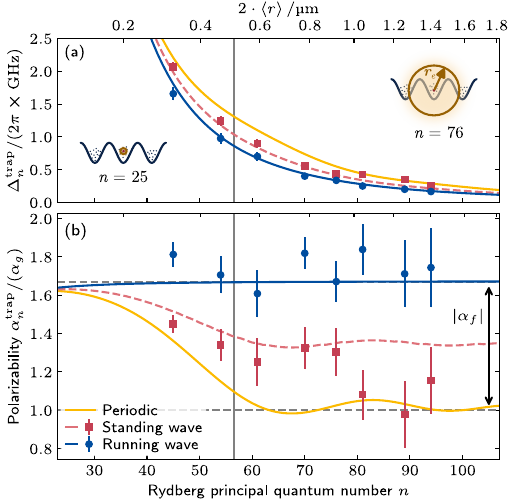}
    \caption{Magic detunings for our running-wave and standing-wave trap as function of Rydberg state. (a) Optimal detunings $\Delta^\mathrm{trap}_{n}$ determined by measurements as in Fig.~\ref{fig4:photon_storage_measurements}(c) for a range of Rydberg principal quantum numbers $n$ and for the different trap configurations.
    The solid lines show the analytic solution for the running wave (blue) and the periodic potential (yellow), from Eqs.~(\ref{eq:magic condition running wave}) and (\ref{eq:magic condition periodic}) respectively.
    The optimal detuning~$\Delta^\mathrm{sw}_{n}$ for the full standing wave is extracted from the standard deviation of the differential light shift across the atomic cloud (red, dashed line).
    The vertical gray line marks the standing-wave trap periodicity ($\SI{506}{\nano\meter}$) in comparison to the Rydberg electron orbital diameter $2\cdot\ev{r}$ on the top x-axis.
    Experimental data for the running wave and the standing wave are shown as blue circles and red squares respectively.
    The two schematic drawings highlight the size of the Rydberg electron wave function in relation to the standing-wave trap periodicity.
    (b) Polarizability calculated from the data shown in (a) as described by Eq.~(\ref{eq:norm polarizability}).
    }
    \label{fig5:polarizability_vs_n}
\end{figure}

We show in Fig.~\ref{fig5:polarizability_vs_n}(a) the optimal detunings $\Delta^\mathrm{trap}_n$ for different principal quantum numbers $n$ and the two trap configurations.
The magic detunings move closer to resonance for higher $n$, because of the smaller dipole matrix elements $D_{ar}\propto (n-3.13)^{-3/2}$.
From the experimental detunings and the known matrix elements we extract an $n$-dependent effective polarizability for our experimental configurations defined as
\begin{equation}
    \alpha^{\mathrm{trap}}_n: = \frac{D_{ar}^2}{4\hbar \Delta^{\mathrm{trap}}_{n}}.
    \label{eq:norm polarizability}
\end{equation}
In Fig.~\ref{fig5:polarizability_vs_n}(b) we show the increasing difference of this polarizability for the two trap geometries as function of principal quantum number.
The running wave polarizability $\alpha^\mathrm{rw}_n$ and the periodic trap polarizability $\alpha^\sim_n$ satisfy the magic conditions
\begin{align}
    \alpha^\mathrm{rw}_n &= \alpha_g - \alpha_f
    \label{eq:norm polarizability condition rw}\\
    \alpha^\sim_n &= \alpha_g - \alpha_f \cdot \theta_n,
    \label{eq:norm polarizability condition sw}
\end{align}
and show in their difference the $n$ scaling of the landscape factor $\theta_n$, see Eq.~(\ref{eq:theta_n approx}).
The difference between running and standing wave indicates that the trapping geometry plays an important role for the optical confinement of high $n$ Rydberg states.
The top axis in Fig.~\ref{fig5:polarizability_vs_n} shows the Rydberg electron orbital diameter $2\cdot\ev{r}$ as a function of Rydberg state.
For low $n$ ($n\lesssim 30$), the size of the Rydberg electron wave function is small in comparison to the lattice periodicity, as shown in Fig.~\ref{fig5:polarizability_vs_n}(a) for $n=25$.
The case for Rydberg state $n = 76$ is shown as well, where the electron orbital already spans more than one lattice site. Here, the magic conditions between running and standing wave are modified significantly.
For very large $n$ ($n\gtrsim 100$) the landscape factor goes to zero as the free electron samples more and more lattice sites. The resulting difference between $\alpha^\mathrm{rw}$ and $\alpha^\sim$ becomes constant, shown by the arrow with length $|\alpha_f|/\alpha_g$.

\section{Conclusion}
\label{sec:conclusion}
In this paper we studied how coherence between ground and Rydberg state depends on the trap geometry. This dependence is caused by the Rydberg electron that, due to the non-negligible size of its wave function, explores the potential landscape around the Rydberg atom. 
To obtain an intuitive picture of this effect, we present simple calculations and graphical visualizations of the differential light shift in running-wave and standing-wave traps.
We experimentally determine the detunings of least dephasing for a range of Rydberg states for both traps. The result shows that the trapping beam geometry plays an important role when creating magic optical traps for Rydberg atoms. The measured detunings are in good agreement with the values expected from our calculations.
Our work shows that it is necessary to take the size of both Rydberg atom and trapping beams into account when trapping Rydberg atoms. The results obtained in this work are of importance for magic trapping of Rydberg polaritons, collectively excited ensembles, and for experiments with individual Rydberg atoms.

\section*{Acknowledgments}
We are grateful to Wilson S. Martins for valuable discussions.
The experimental data and analysis code underlying the results presented in this paper is freely accessible in the open-access Zenodo database under \url{https://doi.org/10.5281/zenodo.14003106}.
The research leading to these results has received funding from the Deutsche Forschungsgemeinschaft (DFG, German Research Foundation) under Project No. 449905436, Research Unit FOR 5413/1, Grant No. 465199066, and Germany’s Excellence Strategy – Cluster of Excellence Matter and Light for Quantum Computing (ML4Q) EXC 2004/1 – 390534769, from the European Union’s Horizon 2020 program under the ERC grant SUPERWAVE (grant No.101071882). IL is grateful for financing from the Baden-Württemberg Stiftung through Project No.~BWST\textunderscore ISF2019-23.

\appendix

\onecolumngrid

\section{Energy correction caused by \texorpdfstring{$H_1$}{H1}}
\label{appendix:H1-correction}
We derive the energy correction of an eigenstate $\ket q$ of the unperturbed Hamiltonian $H_0$ due to $H^{(1)}$ from Eq.~(\ref{eq:general-Hamiltonian}).
We utilize time-dependent perturbation theory,
specifically, the Floquet quasi-energy formalism~\cite{krainov1985,krainov1999}.
The first order correction in $H_1$ vanishes due to parity arguments of the dipole operator.
Thus, we derive the second order energy shift of an atom in state $\ket q$.
To do so, we expand $H_1$ in Floquet-modes $m$ and therefore express the vector potential $\bm{A}(\bm{r},t)$ in factors $e^{im\omega\tau}$, see Eq.~(\ref{eq:A-e-iomegat}).
From this we see that the only contribution is for $m=1$, thus
\begin{align}
    U_q^{(1)}
    &\approx\frac{e^2}{m_\mathrm e^2}\sum_{j\neq q}
    \frac{1}{\hbar}\left(
        \frac{\abs{
        \mel{q}{\bm{A}(\bm{r}+\bm{r}_\mathrm{e})\cdot\bm{p}_\mathrm{e}}{j}}^2}
            {\omega_{qj}-\omega}
        +\frac{\abs{\mel{q}{\bm{A}^*(\bm{r}+\bm{r}_\mathrm{e})\cdot\bm{p}_\mathrm{e}}{j}}^2}{\omega_{qj}+\omega}\right),
\end{align}
where we used $\bm p_\mathrm e\cdot \bm{A}(\bm{r}+\bm{r}_\mathrm e)=\bm{A}(\bm{r}+\bm{r}_\mathrm e)\cdot \bm p_\mathrm e$ in Coulomb gauge.
We now expand the vector potential
$\bm{A}(\bm{r}+\bm{r}_\mathrm{e})=\bm{A}(\bm{r})+\bm\nabla_{\bm{r}}\bm{A}(\bm{r})\cdot\bm{r}_\mathrm e+\dots$ for a small electron radius $\bm{r}_\mathrm e$.
We use $\bm A(\bm r)=\bm \epsilon A(\bm r)$ and the commutation identity of the hydrogen atom
$\mel{q}{\bm{\epsilon}\cdot \bm{p}_\mathrm{e}}{j}=im_\mathrm e \omega_{qj} \mel{q}{e\,\bm{\epsilon}\cdot\bm{r}_\mathrm{e}}{j}$~\cite{Traini1996} to simplify the expectation value with
\begin{align}
    U_q^{(1)}
    =& \frac{e^2}{m^2_\mathrm{e}}\sum_{j\neq q}
    \frac{1}{\hbar}\left(
    \frac{\abs{
        \mel**{q}{\left[\bm{A}(\bm{r})+\bm\nabla_{\bm{r}}{\bm{A}(\bm{r})}\cdot\bm{r}_\mathrm{e}+\dots\right]\cdot\bm{p}_\mathrm{e}}{j}}^2}{\omega_{qj}-\omega}
    +\frac{\abs{
        \mel**{q}{\left[\bm A^*(\bm r)+\bm\nabla_{\bm{r}}{\bm{A}^*(\bm{r})}\cdot\bm{r}_\mathrm e+\dots\right]\cdot\bm{p}_\mathrm{e}}{j}}^2}{\omega_{qj}+\omega}\right)\\
    =& \frac{e^2}{m^2_\mathrm{e}}\sum_{j\neq q}
     \frac{1}{\hbar}\left(
     \frac{\Big|
        A(\bm r)\mel**{q}{\bm \epsilon\cdot\bm{p}_\mathrm{e}}{j}
       +\mel**{q}{[\bm\nabla_{\bm{r}}\bm{A}(\bm{r}) \cdot \bm{r}_\mathrm e] \cdot\bm{p}_\mathrm{e}}{j}
       +\dots
     \Big|^2}{\omega_{qj}-\omega}
     +\frac{\Big|
        A^*(\bm r)\mel**{q}{\bm \epsilon\cdot\bm{p}_\mathrm{e}}{j}
       +\mel**{q}{[\bm\nabla_{\bm{r}}\bm{A}^*(\bm{r}) \cdot \bm{r}_\mathrm{e}] \cdot\bm{p}_\mathrm{e}}{j}
       +\dots
     \Big|^2}{\omega_{qj}+\omega}\right)\\
     =&\frac{e^2}{\hbar m^2_\mathrm e}\sum_{j\neq q}\left[
     \abs{A(\bm r)}^2\Bigg(
     \frac{ \abs{   \mel**{q}{\bm \epsilon\cdot\bm{p}_\mathrm{e}}{j}}^2}      {\omega_{qj}-\omega}
     +\frac{ \abs{   \mel**{q}{\bm \epsilon^*\cdot\bm{p}_\mathrm{e}}{j}}^2}      {\omega_{qj}+\omega}
     \Bigg)
     +
    \left( \frac{\abs{\mel**{q}{[\bm\nabla_{\bm{r}}\bm A(\bm{r}) \cdot \bm{r}_\mathrm e] \cdot\bm{p}_\mathrm{e}}{j}}^2}{\omega_{qj}-\omega}
        +\frac{\abs{\mel**{q}{[\bm\nabla_{\bm{r}}\bm A^*(\bm{r}) \cdot \bm{r}_\mathrm e] \cdot\bm{p}_\mathrm{e}}{j}}^2}{\omega_{qj}+\omega}
        \right)\right]+\dots
    \\
     =&-\abs{\bm A(\bm r)}^2\omega^2
     \underbrace{\left[-\sum_{j\neq q} \frac{\omega_{qj}^2}{\omega^2}\Bigg(
     \frac{ \abs{   \mel**{q}{e\bm \epsilon\cdot\bm{r}_\mathrm{e}}{j}}^2}{\hbar(\omega_{qj}-\omega)}
     +\frac{ \abs{   \mel**{q}{e\bm \epsilon^*\cdot\bm{r}_\mathrm{e}}{j}}^2}{\hbar(\omega_{qj}+\omega)}
     \Bigg)\right]}_{\text{bound-state polarizability }\alpha_\mathrm q^{(\mathrm b)}}\nonumber\\
     &+\frac{e^2}{\hbar m_\mathrm{e}^2}\sum_{j\neq q}
    \left( \frac{\abs{\mel**{q}{[\bm\nabla_{\bm{r}}\bm A(\bm{r}) \cdot \bm{r}_\mathrm e] \cdot\bm{p}_\mathrm{e}}{j}}^2}{\omega_{qj}-\omega}
        +\frac{\abs{\mel**{q}{[\bm\nabla_{\bm{r}}\bm A^*(\bm{r}) \cdot \bm{r}_\mathrm e] \cdot\bm{p}_\mathrm{e}}{j}}^2}{\omega_{qj}+\omega}
        \right)+\dots
    \label{eq:hyperpolarizability-correction}
\end{align}
Here, the first summand is the dynamic dipole polarizability $\alpha_q^{(\mathrm{b})}(\omega)$ of the bound states in the atom.
Note that $\alpha_q^{(\mathrm{b})}(\omega)$ is not the dynamic polarizability of the atom in state $\ket{q}$, since there is an additional contribution from the coupling to the continuum of unbound states \cite{jones2021}.
When setting $\ket{q}=\ket{g}$, we obtain the bound state contribution of the dynamic polarizability of the ground state atom $\alpha_g^{(b)}$, see Eq.~(\ref{eq:U-g-1}) of the main text.

The second summand in Eq.~(\ref{eq:hyperpolarizability-correction}) represents the quadrupole polarizability arising from the gradient of the vector potential~\cite{krainov1985}.
For a ground state atom this contribution is negligible, since the gradient is negligible over the size $r_\mathrm{e}$ of the atom.
The contributions from higher order expansions of the potential $\bm{A}(\bm{r}+\bm{r}_\mathrm{e})$ lead to higher order polarizabilities (i.e. hyperpolarizabilities). 
The first higher order term contains a quadrupole transition between $\ket{g}$ and $\ket{j}$, the second an octupole transition, and so on.
These contributions are negligible compared to the dipole transition, and consequently, the energy correction for a ground state atom due to $H^{(1)}$ reduces to Eq.~(\ref{eq:U-g-1}) with
\begin{equation}
     U_g^{(1)}(\bm{r})\approx-\alpha_g^{(\mathrm{b})} \omega^2{\abs{\bm{A}(\bm{r})}^2}.
\end{equation}

If the atom is in the Rydberg state ($\ket{q}=\ket{r}$), the calculation of the polarizability includes the summation over many matrix elements.
However, the frequency $\omega$ is chosen to be near resonant to the hyperfine state $\ket{a}=\ket{6P_{3/2},F=3,m_F=3}$.
In the experiment, the trap laser has $\sigma^-$ polarization. For this polarization, all other hyperfine transitions to $6P_{3/2}$ are dipole forbidden.
Therefore, we can perform the resonance approximation, since terms including other transitions than $\ket{a}$ to $\ket{r}$ are smaller by orders of magnitude.

The contribution from the quadrupole transitions in Eq.~(\ref{eq:hyperpolarizability-correction}) is strongly suppressed since the trapping laser is off-resonant to these transitions. Additionally, our trap laser couples the Rydberg state to a low-lying state. Thus, the expectation value is primarily influenced by the spatial extent of the low-lying state’s wave function, which defines the region over which the electric field gradient is evaluated, and the gradient is small in this region. Therefore, we only consider the contribution from the resonant dipole transition.

With these approximations, $U_r^{(1)}$ simplifies to
\begin{align}
    U_r^{(1)}
    &\approx\abs{\bm{A}(\bm{r})}^2
    \frac{\omega_{ra}^2}{\hbar}\left(
    \frac{\abs{\mel{r}{e\,\bm \epsilon \cdot\bm{r}_\mathrm{e}}{a}}^2}{\omega_{ra}-\omega}
    +\frac{\abs{\mel{r}{e\,\bm \epsilon^* \cdot\bm{r}_\mathrm{e}}{a}}^2}{\omega_{ra}+\omega}\right).
\end{align}

We now perform the rotating wave approximation with $\omega_{ra}+\omega\gg\omega_{ra}-\omega$ and introduce the detuning $\Delta=\omega-\omega_{ra}$. We use $\omega_{ra}^2\approx\omega^2$ and obtain
\begin{align}
    U_r^{(1)}
    &\approx\omega^2\abs{\bm{A}(\bm{r})}^2
    \frac{\abs{\mel{r}{e\,\bm \epsilon \cdot\bm{r}_\mathrm{e}}{a}}^2}{\hbar\Delta}.
\end{align}

We express the dipole matrix elements in terms of the reduced dipole matrix element $D_{ar}$ for our polarization~$\bm \epsilon$ as
\begin{align}
    \abs{\mel{r}{e\,\bm \epsilon \cdot\bm{r}_\mathrm{e}}{a}}^2
    &=\abs{\mel**{r}{\frac{e}{\sqrt 2}(\bm{\hat{e}}_x +i\bm{\hat{e}}_y) \cdot\bm{r}_\mathrm{e}}{a}}^2
    =\frac{1}{2}\abs{\mel**{r}{e(\bm x_\mathrm{e} +i\bm y_\mathrm{e}) }{a}}^2\\
    &\eqdef \frac{1}{2}\abs{\frac{1}{2}\langle nS_{1/2},J=1/2||d||6P_{3/2},J=3/2\rangle}^2\\
    &\eqdef \frac{1}{8} D_{ar}^2.
\end{align}
We calculate the reduced dipole element of this transition using the ARC package~\cite{jones2021}.
By combining these results with the definition of the vector potential $\bm{A}(\bm{r})$, we arrive at Eq.~(\ref{eq:U_r1}) in the main text, which we write out as
\begin{align}
    U_r^{(1)}
    &\approx-\frac{D_{ar}^2}{8\hbar\Delta}
    \omega^2{\abs{\bm{A}(\bm{r})}^2}
    = -\frac{D_{ar}^2}{16\hbar\Delta}\left[\left(E_\rightarrow(\bm{r})
        -E_\leftarrow (\bm{r})\right)^2
        +4E_\rightarrow(\bm{r}) E_\leftarrow (\bm{r})\cos(\bm k\bm{r})\right]
    .
\end{align}

\section{Vector potential}
\label{appendix:vector-potential}
The vector potential of Eq.~(\ref{eq:vector-potential}) can be reformulated using trigonometric relations.
Note that we assume that the field amplitude is homogeneous over the diameter of the Rydberg (ground state) atom. Thus, we set $E_{\rightarrow(\leftarrow)}(\bm{r}+\bm{r}_\mathrm{e})\approx E_{\rightarrow(\leftarrow)}(\bm{r})$.
We then write
\begin{alignat}{4}
&\omega\bm{A}(\bm{r}+\bm{r}_\mathrm e,t)
    &&=&&&&\sin{\left(\bm k \bm{r}_\mathrm e \right)}\bm\epsilon\Bigg[
        E_\leftarrow(\bm{r})\left(
            \sin{\left(\bm k \bm{r}  \right)} \sin{\left(\omega t \right)}
            -\cos{\left(\bm k \bm{r}  \right)} \cos{\left(\omega t \right)}
        \right)
        +E_\rightarrow(\bm{r})\left(
            \sin{\left(\bm k \bm{r}  \right)} \sin{\left(\omega t \right)}
            +\cos{\left(\bm k \bm{r}  \right)} \cos{\left(\omega t \right)}
        \right)
    \Bigg]\nonumber\\
    &&&&&+&&\cos{\left(\bm k \bm{r}_\mathrm e \right)}\bm\epsilon\Bigg[
        E_\leftarrow(\bm{r})\left(
            - \sin{\left(\bm k \bm{r}  \right)} \cos{\left(\omega t \right)}
            - \cos{\left(\bm k \bm{r}  \right)} \sin{\left(\omega t \right)}
        \right)
        + E_\rightarrow(\bm{r})\left(
            \sin{\left(\bm k \bm{r}  \right)} \cos{\left(\omega t \right)}
            -\cos{\left(\bm k \bm{r}  \right)} \sin{\left(\omega t \right)}
            \right)
        \Bigg].
\end{alignat}
The dependence of the electron radius $\bm{r}_\mathrm e$ is factored out now.
Alternatively, the vector potential can be expressed as
\begin{align}
    \bm{A}(\bm{r}+\bm{r}_\mathrm e,t)=
    &\underbrace{\bm\epsilon\left( \frac{(E_\rightarrow(\bm{r})-E_\leftarrow(\bm{r}))}{2w}\sin{(\bm k(\bm{r}+\bm{r}_\mathrm e))} -\frac{i (E_\rightarrow(\bm{r})+E_\leftarrow(\bm{r}))}{2w}\cos{(\bm k(\bm{r}+\bm{r}_\mathrm e))}\right)}_{\bm{A}(\bm{r}+\bm{r}_\mathrm e)} e^{- i w t}
    + \bm{A}^{*}(\bm{r}+\bm{r}_\mathrm e) e^{i w t}
,
    \label{eq:A-e-iomegat}
\end{align}
where $\bm{A}(\bm{r}+\bm{r}_\mathrm e)$ is the time-independent vector potential.
This representation is needed for the time dependent perturbation theory using Floquet quasi-energies.
It follows,
\begin{align}
    \bm{A}^2(\bm{r}+\bm{r}_\mathrm e,t)=
    \bm{A}^2(\bm{r}+\bm{r}_\mathrm e)e^{-2i\omega t}
    +2\bm{A}(\bm{r}+\bm{r}_\mathrm e) \bm{A}^*(\bm{r}+\bm{r}_\mathrm e)
    +(\bm{A}^*)^2(\bm{r}+\bm{r}_\mathrm e)e^{2i\omega t}.
    \label{eq:A2-e-2iomegat}
\end{align}
For the time-independent squared vector potential we obtain
\begin{align}
4\omega^2{\abs{\bm{A} (\bm{r}+\bm{r}_\mathrm e)}}^2
    = &E_\rightarrow(\bm{r})^2+E_\leftarrow(\bm{r})^2
    + 2E_\rightarrow(\bm{r})E_\leftarrow(\bm{r}) \cos(2\bm k \bm{r}_\mathrm e) \left(2 \cos^2(\bm k \bm{r} ) - 1\right)\nonumber\\
    &- 4 E_\rightarrow(\bm{r}) E_\leftarrow(\bm{r}) \sin(2\bm k \bm{r}_\mathrm e) \sin{\left(\bm k \bm{r} \right)} \cos{(\bm k\bm{r})}
.
    \label{eq:vector-potential-avg-squared}
\end{align}
In the limit of $\bm k\bm{r}_\mathrm e\rightarrow 0$, this expression reduces to
\begin{align}
4\omega^2{\abs{\bm{A}(\bm{r})}}^2
&=  4E_\rightarrow (\bm{r})E_\leftarrow (\bm{r})\cos^2(\bm k \bm{r}) + \left(E_\rightarrow (\bm{r})- E_\leftarrow (\bm{r}) \right)^{2}.
\label{eq:vector-potential-avg-dipole-squared}
\end{align}

\twocolumngrid
\bibliography{references}

\end{document}